\documentclass[a4paper,12pt]{article}

\usepackage{amsmath}
\usepackage{amssymb}
\usepackage[numbers, square, comma, sort&compress]{natbib}
\usepackage{hyperref}
\usepackage{graphicx}
\usepackage{epstopdf}
\usepackage{bm}
\usepackage{epsfig}
\usepackage{graphics}
\usepackage{xspace}
\usepackage{amsfonts}
\usepackage{verbatim}
\usepackage{geometry}
\usepackage{color}
\allowdisplaybreaks

\newcommand{\nn}{\nonumber}

\newcommand{\beq}{\begin{equation}}
\newcommand{\eeq}{\end{equation}}
\newcommand{\bqa}{\begin{eqnarray}}
\newcommand{\eqa}{\end{eqnarray}}

\def\red[#1]{{\color{red}#1}}

\begin{document}
%%%%%%%%%%%%%%%%%%%%%%%%%%%%%%%%%%%%%%%%%%
%Define Title, Author, Address, Preprint#

\begin{titlepage}
\begin{flushright}
TUM-HEP-1157/18\\ [0.2cm]
\today
\\
\end{flushright}

\vskip 25mm

\begin{center}
\Large\bf{Three-loop  planar  master integrals for heavy-to-light form factors }
\end{center}

\vskip 8mm

\begin{center}

{\bf Long-Bin Chen$^{a}$, Jian Wang$^b$}\\
\vspace{10mm}
\textit{$^a$School of Physics and Electronic Engineering, Guangzhou University, Guangzhou 510006, China}\vspace{2mm}\\
\textit{$^b$Physik Department T31,  Technische Universit\"at M\"unchen, James-Franck-Stra\ss e~1,
D--85748 Garching, Germany} \\
\vspace{5mm}
\textit{E-mail:} \texttt{{\href{chenlb@gzhu.edu.cn}{chenlb@gzhu.edu.cn},
\href{j.wang@tum.de}{j.wang@tum.de}}}

\end{center}

\vspace{10mm}

%%%%%%%%%%%%%%%%%%%%%%%%%%%%%%%%%%%%%%%%%%
%Create the title page

\begin{abstract}
We calculate analytically the three-loop  planar master integrals
relevant for heavy-to-light form factors  using the method of differential equations.
After choosing a proper canonical basis, the boundary conditions are easy to be determined,
and the solution of differential equations is greatly simplified.
The results for seventy-one master integrals at general kinematics are all expressed in terms of harmonic polylogarithms.
\end{abstract}

\end{titlepage}

%%%%%%%%%%%%%%%%%%%%%%%%%%%%%%%%%%%%%%%%%%
%Main body of the paper

\section{Introduction}

Huge samples of top quarks at the LHC and $B$ mesons in $B$-factories provide us good opportunity
to precisely measure the properties of heavy quarks,
e.g., the  CKM matrix element $|V_{ub}|$, which may give some clues for the new physics.
To match the increasing experimental precision, higher-order theoretical predictions are mandatory.

The heavy quark form factor is an important ingredient of the higher-order calculations on the heavy quark production,
and has been explored up to the third order in $\alpha_s$ \cite{Henn:2016tyf,Henn:2016kjz,Lee:2018rgs,Ablinger:2018yae}.
In contrast, the heavy-to-light form factor, which is a base for the calculation of the heavy quark or meson decay \cite{Chetyrkin:1999ju,Blokland:2004ye,Czarnecki:2005vr,Bonciani:2008wf,Asatrian:2008uk,Beneke:2008ei,Bell:2008ws,Gao:2012ja,Brucherseifer:2013iv},
has been studied so far only up to the second order \cite{Bonciani:2008wf,Huber:2009se,Bell:2006tz}.
In order to achieve a uniform precision for the production and decay, it is necessary to
improve the knowledge of the heavy-to-light form factor to the third order too.
In this work, we calculate, as a first step toward this goal,  the three-loop planar master integrals for the heavy-to-light form factors.
We consider an arbitrarily momentum transfer so that our result can be applied not only in a heavy quark decay,
but also in other processes, such as $W'\to t\bar{b}$ or heavy quark production via deep inelastic scattering.

Another motivation of our work is the understanding of the infrared divergences of the amplitude involving both massless and massive particles.
The general structure of the infrared divergence in a massive amplitude has been investigated up to two loop level
\cite{Kidonakis:2009ev,Mitov:2009sv,Becher:2009kw,Beneke:2009rj,Ferroglia:2009ep,Ferroglia:2009ii,Mitov:2010xw}.
It is interesting to explore the structure at even higher orders.
Our project, once completed, would help to see this infrared structure.

Our calculation relies on the method of differential equations \cite{Kotikov:1990kg,Kotikov:1991pm},
and is highly inspired by the strategy of choosing a proper basis with uniform transcendentality proposed in ref.\cite{Henn:2013pwa}.
This choice simplifies the solution of differential equations significantly,
and has been successfully applied in \cite{Henn:2013woa,Henn:2013fah,Henn:2014lfa,Gehrmann:2014bfa,Caola:2014lpa,DiVita:2014pza,Huber:2015bva,Bonciani:2016ypc,Mastrolia:2017pfy,Chen:2015csa,Chen:2017xqd,Chen:2018dpt} and many other works.
In addition, we utilize several properties of the master integrals around the (pseudo-)singularities
to reduce the determination of the boundary condition
to simple  or even trivial  integrals.

This paper is organized as follows. In section \ref{sec2} we present the canonical basis and the corresponding differential equations.
We discuss the determination of boundary conditions in section \ref{sec3}.
Conclusions are given in section \ref{sec4}.
The analytic results as well as the rational matrices are provided in ancillary files.

\section{Canonical basis and differential equations}
\label{sec2}

\begin{figure}[ht]
\begin{center}
\includegraphics[scale=0.34]{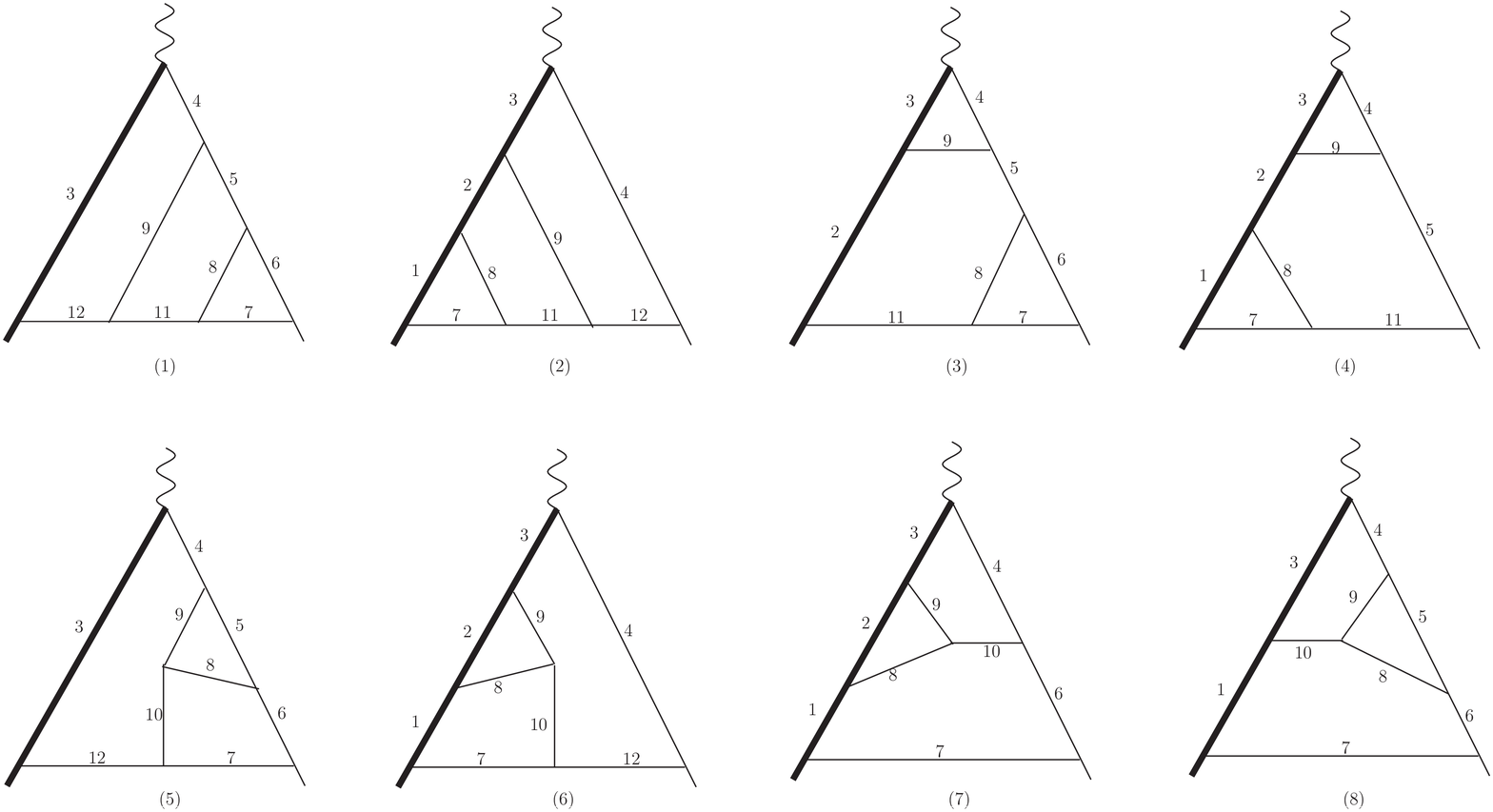}\\
\includegraphics[scale=0.34]{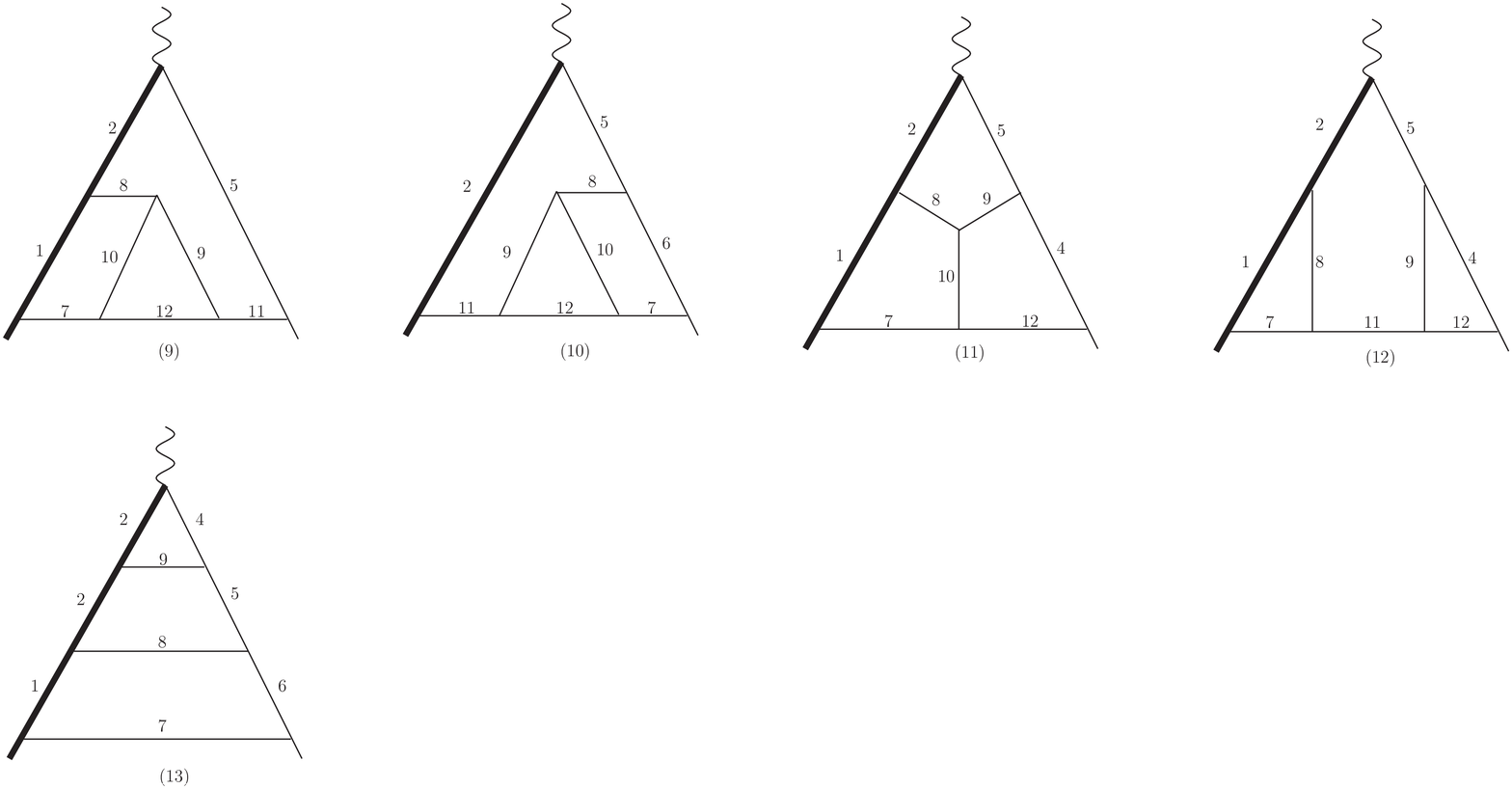}
\caption{The color-planar three-loop topologies for the heavy-to-light form factor.
The bold lines represent massive particles, while the thick lines indicate massless particles.
}
\label{midiag1}
\end{center}
\end{figure}

Though our goal is to calculate the full three-loop heavy-to-light form factor,
we focus first on the leading color contribution, which constitutes a gauge invariant part and dominates the full result
in the large $N_c$ limit.
As such, we consider the color-planar integrals in this work.
The corresponding topologies of the integral families are shown in figure \ref{midiag1}.
All these scalar integrals can be  formulated as
\bqa
I_{n_1,n_2,\ldots,n_{12}}=\int\frac{{\mathcal D}^d k_1~{\mathcal D}^d k_2~{\mathcal D}^d k_3}{D_1^{n_1}~D_2^{n_2}~D_3^{n_3}~D_4^{n_4}~D_5^{n_5}~D_6^{n_6}~D_7^{n_7}~D_8^{n_8}~D_9^{n_9}~ D_{10}^{n_{10}}~D_{11}^{n_{11}}~D_{12}^{n_{12}}}
\label{def}
\eqa
with the propagator defined by
\begin{align}
D_1&=-(k_1+p_1)^2+m^2,D_2=-(k_2+p_1)^2+m^2,D_3=-(k_3+p_1)^2+m^2,\nonumber\\
D_4&=-(k_3+p_2)^2,D_5=-(k_2+p_2)^2,D_6=-(k_1+p_2)^2,D_7=-k_1^2,\nonumber\\
D_8&=-(k_1-k_2)^2,D_9=-(k_2-k_3)^2,D_{10}=-(k_1-k_3)^2,D_{11}=-k_2^2,D_{12}=-k_3^2,\nonumber
\label{int}
\end{align}
and
\beq
{\mathcal D}^d k_i \equiv \frac{m^{2\epsilon}}{\pi^{d/2}\Gamma(1+\epsilon)}  d^d k_i,\quad d=4-2\epsilon \ .
\eeq
The heavy and light quarks are both on-shell, i.e., $p_1^2=m^2,p_2^2=0$, and we consider an arbitrary momentum transfer $(p_1-p_2)^2\equiv s$.
All the color-planar scalar integrals considered can be described by the parametrization of eq. (\ref{def}). Note that for the definition of planar integrals in eq. (\ref{def}), each index can be positive but the total number of positive indices is less than or equal to nine.

The first step of the calculation is to set up the differential equations.
The derivative with respect to the kinematic variable $s$ can be written in terms of a derivative respect to the external momentum,
\bqa
\frac{\partial}{\partial\, s}=\frac{1}{s-m^2}p_2 \cdot\frac{\partial}{\partial \, p_2}.
\eqa

The planar scalar integrals shown in figure~\ref{midiag1} can be reduced to a set of 71 master integrals by using the \texttt{FIRE} package \cite{Smirnov:2014hma}.
Inspired by the way of choosing canonical basis presented in \cite{Henn:2016kjz,Henn:2013fah}, we choose a basis ${\bf F}=\{F_1,\ldots,F_{71}\}$ for the master integrals,
\begin{align}
F_1 &= m^6\, I_{3,3,3,0,0,0,0,0,0,0,0,0}\, ,\nn\\
F_2 &= \epsilon^2\, m^4\, I_{0, 2, 3, 0, 0, 0, 1, 2, 0, 0, 0, 0} \, ,\nn\\
F_3 &= \epsilon^3\, m^2\, I_{0, 0, 2, 0, 0, 0, 2, 2, 1, 0, 0, 0} \, ,\nn\\
F_4 &= (\epsilon-1)(1+4\epsilon)\epsilon\, m^2\, I_{2, 0, 2, 0, 0, 0, 0, 2, 1, 0, 0, 0} \, ,\nn\\
F_5 &= \epsilon\, s\, m^4\, I_{3, 3, 2, 1, 0, 0, 0, 0, 0, 0, 0, 0} \, ,\nn\\
F_6 &= \epsilon^3\, s\, I_{2, 0, 0, 2, 0, 0, 0, 2, 1, 0, 0, 0} \, ,\nn\\
F_7 &= \epsilon^2\, m^2\, (2\epsilon\, I_{2, 0, 0, 2, 0, 0, 0, 2, 1, 0, 0, 0} + (s-m^2)\, I_{3, 0, 0, 2, 0, 0, 0, 2, 1, 0, 0, 0}) \, ,\nn\\
F_8 &= \epsilon^2\, s\, m^2\, I_{2, 0, 3, 0, 2, 0, 0, 1, 0, 0, 0, 0} \, ,\nn\\
F_9 &= \epsilon^2\, m^2\, (-2m^2\, I_{2, 0, 3, 0, 2, 0, 0, 1, 0, 0, 0, 0} + (s-m^2)\, I_{1, 0, 3, 0, 2, 0, 0, 2, 0, 0, 0, 0}) \, ,\nn\\
F_{10} &= \epsilon^2\, s^2\, m^2\, I_{3, 2, 2, 1, 1, 0, 0, 0, 0, 0, 0, 0} \, ,\nn\\
F_{11} &= \epsilon^4\, (s-m^2)\, I_{0, 0, 1, 1, 0, 0, 2, 2, 1, 0, 0, 0} \, ,\nn\\
F_{12} &= \epsilon^3\, m^2\, I_{0, 2, 0, 0, 0, 0, 2, 1, 2, 0, -1, 1} \, ,\nn\\
F_{13} &= \epsilon^4\, (s-m^2)\, I_{0, 2, 0, 2, 0, 0, 1, 1, 0, 1, 0, 0} \, ,\nn\\
F_{14} &= \epsilon^3\, m^2\, (I_{2, -1, 2, 0, 0, 0, 0, 1, 1, 0, 2, 0}- m^2\, I_{2, 0, 2, 0, 0, 0, 0, 1, 1, 0, 2, 0}) \, ,\nn\\
F_{15} &= \epsilon^3\, m^2\, (s-m^2)\, I_{0, 1, 3, 0, 1, 0, 2, 1, 0, 0, 0, 0}-\frac{\epsilon^2}{2}\, m^4\, I_{0, 2, 3, 0, 0, 0, 1, 2, 0, 0, 0, 0} \, ,\nn\\
F_{16} &= \epsilon^3\, s\, m^2\, I_{0, 2, 2, 1, 0, 0, 1, 2, 0, 0, 0, 0} \, ,\nn\\
F_{17} &= \epsilon^3(1-2\epsilon)\, \frac{s}{s-m^2}\, I_{2, 0, 1, 1, 0, 0, 0, 1, 2, 0, 0, 0}
- \frac{s}{6(s-m^2)}(F_4+2F_7)-\frac{4s-3m^2}{3(s-m^2)}F_6 \, ,\nn\\
F_{18} &= \epsilon^4\, (s-m^2)\, I_{0, 1, 0, 2, 0, 0, 2, 1, 1, 0, 0, 0} \, ,\nn\\
F_{19} &= \epsilon^3(1-4\epsilon)\, \frac{s}{s+m^2}\, I_{0, 1, 0, 2, 0, 0, 2, 1, 1, 0, -1, 0}- \, \frac{5 m^2}{s+m^2}\, F_{18} \, ,\nn\\
F_{20} &= \epsilon^4(1-2\epsilon)\, s\, I_{0, 2, 0, 1, 0, 1, 0, 1, 1, 1, 0, 0} \, ,\nn\\
F_{21} &= -\epsilon^3(1+2\epsilon)(s-m^2)I_{0, 1, 0, 2, 0, 2, 0, 1, 1, 0, 0, 0}-\frac{s-m^2}{s}F_{20} \nn\\
       &+\left(\frac{1}{3}+\frac{m^2}{s}\right)F_6-\frac{2}{3}F_7,\nn\\
F_{22} &= \epsilon^4\, (s-m^2)\, I_{2, 0, 0, 0, 1, 0, 0, 1, 2, 0, 0, 1} \, ,\nn\\
F_{23} &= \epsilon^3\, m^2\, (s-m^2)\, I_{3, 0, 0, 0, 1, 0, 0, 1, 2, 0, 0, 1} \, ,\nn\\
F_{24} &= \epsilon^4(1-2\epsilon)\, s\, I_{1, 0, 1, 0, 2, 0, 0, 1, 1, 1, 0, 0} \, ,\nn\\
F_{25} &= \epsilon^3(1-2\epsilon)\, \frac{s\, m^2\, (s-m^2)}{3s-m^2}\, I_{2, 0, 1, 0, 2, 0, 0, 1, 1, 1, 0, 0}+\frac{8 \, m^2}{3(3s-m^2)}F_{24} \, \nn\\
&- \frac{s}{2(3s-m^2)} (F_4-2F_7+4F_9) +\frac{2 m^2}{3(3s-m^2)}(F_6+3F_8)\, , \nn\\
F_{26} &= \epsilon^3 \, s^2 \, I_{2, 0, 2, 1, 2, 0, 0, 1, 0, 0, 0, 0} \, , \nn\\
F_{27} &= \epsilon^3\, s \,  (s-m^2) I_{1, 0, 2, 1, 2, 0, 0, 2, 0, 0, 0, 0}-\frac{2 m^2}{s} \, F_{26} \, ,\nn\\
F_{28} &= \epsilon^4 \, (s-m^2) \, I_{0, 1, 0, 0, 1, 0, 1, 2, 2, 0, -1, 1} \, , \nn\\
F_{29} &= \epsilon^5 \, (s-m^2) \, I_{0, 1, 0, 1, 1, 0, 1, 1, 0, 2, 0, 0} \, , \nn\\
F_{30} &= \epsilon^4 \, s\, (s-m^2) \, I_{0, 1, 2, 1, 1, 0, 1, 2, 0, 0, 0, 0} \, , \nn\\
F_{31} &= \epsilon^4(1-2\epsilon) \, s \, I_{1, 0, 1, 1, 0, 1, 0, 1, 2, 0, 0, 0} \, , \nn\\
F_{32} &= \epsilon^3 \, s^3 \, I_{2, 2, 2, 1, 1, 1, 0, 0, 0, 0, 0, 0} \, , \nn\\
F_{33} &= \epsilon^5 \, (s-m^2) \, I_{0, 2, 1, 1, 0, 0, 1, 1, 0, 1, 0, 0} \, , \nn\\
F_{34} &= \epsilon^4 \, m^2\, (s-m^2) \, I_{0, 3, 1, 1, 0, 0, 1, 1, 0, 1, 0, 0} \, , \nn\\
F_{35} &= \epsilon^4 \, (s-m^2)^2 \, I_{0, 2, 1, 2, 0, 0, 1, 1, 0, 1, 0, 0} \, , \nn\\
F_{36} &= \epsilon^3 \, m^2\, (s-m^2)^2 \, I_{0, 3, 1, 2, 0, 0, 1, 1, 0, 1, 0, 0} \, , \nn\\
F_{37} &= \epsilon^4 \, m^2\, (s-m^2) \, I_{0, 2, 1, 1, 0, 0, 2, 1, 0, 1, 0, 0} \, , \nn\\
F_{38} &= \epsilon^5 \, (s-m^2) \, I_{0, 1, 1, 1, 0, 0, 1, 2, 1, 0, 0, 0} \, , \nn\\
F_{39} &= \epsilon^4 \, m^2\, (s-m^2) \, I_{0, 1, 2, 1, 0, 0, 1, 2, 1, 0, 0, 0} \, , \nn\\
F_{40} &= \epsilon^5 \, (s-m^2) \, I_{2, 0, 0, 0, 1, 0, 0, 1, 1, 1, 0, 1} \, , \nn\\
F_{41} &= \epsilon^4 \, (s-m^2)^2 \, I_{2, 0, 0, 0, 2, 0, 0, 1, 1, 1, 0, 1} \, , \nn\\
F_{42} &= \epsilon^5 \, (s-m^2)^2 \, I_{1, 0, 2, 0, 1, 1, 1, 1, 1, 0, 0, 0} \, , \nn\\
F_{43} &= \epsilon^5 \, (s-m^2) \, I_{1, 0, 2, 0, 1, 0, 1, 1, 1, 0, 0, 0} \, , \nn\\
F_{44} &= \epsilon^4 \, s\, (s-m^2) \, I_{1, 0, 2, 0, 2, 0, 1, 1, 1, 0, 0, 0} \, , \nn\\
F_{45} &= \epsilon^4 \, (s-m^2) \, I_{1, -1, 2, 0, 2, 0, 1, 1, 1, 0, 0, 0}-\frac{m^2}{s}\, F_{44} \, , \nn\\
F_{46} &= \epsilon^5 \, (s-m^2) \, I_{1, 0, 1, 1, 0, 0, 1, 2, 1, 0, 0, 0} \, , \nn\\
F_{47} &= \epsilon^4 \, m^2\, (s-m^2)^2 \, I_{1, 0, 2, 1, 0, 0, 1, 2, 1, 0, 0, 0} \, , \nn\\
F_{48} &= \epsilon^5 \, (s-m^2) \, I_{1, 1, 0, 2, 0, 0, 1, 1, 1, 0, 0, 0} \, , \nn\\
F_{49} &= \epsilon^4 \, m^2\, (s-m^2)^2 \, I_{1, 2, 0, 2, 0, 0, 1, 1, 1, 0, 0, 0} \, , \nn\\
F_{50} &= \epsilon^4 \, m^2\, (s-m^2) \, I_{1, 1, 3, 0, 1, 0, 1, 1, 0, 0, 0, 0} \, , \nn \\
F_{51} &= \epsilon^3 \, m^2\, (s-m^2)^2 \, I_{1, 1, 3, 0, 2, 0, 1, 1, 0, 0, 0, 0} \, , \nn\\
F_{52} &= \epsilon^6 \, (s-m^2) \, I_{0, 1, 1, 1, 0, 1, 1, 1, 1, 0, 0, 0} \, , \nn\\
F_{53} &= \epsilon^5 \, (s-m^2)^2 \, I_{0, 1, 1, 2, 0, 1, 1, 1, 1, 0, 0, 0} \, , \nn\\
F_{54} &= \epsilon^6 \, (s-m^2) \, I_{1, 0, 1, 1, 1, 0, 1, 1, 1, 0, 0, 0} \, , \nn\\
F_{55} &= \epsilon^5 \, (s-m^2)^2 \, I_{1, 0, 1, 1, 1, 0, 1, 2, 1, 0, 0, 0} \, , \nn\\
F_{56} &= \epsilon^5 \, (s-m^2) \, I_{1, 1, 0, 0, 1, 0, 1, 1, 1, 0, -1, 2} \, , \nn\\
F_{57} &= \epsilon^4 \, (-1-6\epsilon)\frac{m^2\, (s-m^2)^2}{s+m^2} \, I_{1, 1, 0, 0, 1, 0, 1, 1, 1, 0, 0, 2}+\frac{3m^2}{s+m^2}F_{56} \,  \nn\\
& -  \frac{m^2}{2(s+m^2)}(2F_2+4F_{15}-4F_{22}+F_{23}+2F_{28}) \, , \nn\\
F_{58} &= \epsilon^6 \, (s-m^2) \, I_{1, 1, 1, 1, 0, 0, 1, 1, 1, 0, 0, 0} \, , \nn\\
F_{59} &= \epsilon^6 \, (s-m^2)^2 \, I_{1, 1, 1, 1, 1, 0, 1, 1, 1, 0, 0, 0} \, , \nn\\
F_{60} &= \epsilon^6 \, (s-m^2)^2 \, I_{1, 1, 1, 1, 0, 1, 1, 1, 1, 0, 0, 0} \, , \nn\\
F_{61} &= \epsilon^6 \, (s-m^2)^3 \, I_{1, 1, 1, 1, 1, 1, 1, 1, 1, 0, 0, 0} \, , \nn\\
F_{62} &= \epsilon^5(1-2\epsilon) \, s \, I_{1, 1, 1, 1, 0, 1, 0, 1, 1, 0, 0, 0} \, , \nn\\
F_{63} &= \epsilon^4(1-2\epsilon) \, s\, m^2 \, I_{2, 1, 1, 1, 0, 1, 0, 1, 1, 0, 0, 0} \, , \nn\\
F_{64} &= \epsilon^5 \, s\, (s-m^2) \, I_{1, 1, 2, 1, 1, 0, 1, 1, 0, 0, 0, 0} \, , \nn\\
F_{65} &= \epsilon^4 \, s\, (s-m^2)^2 \, I_{1, 1, 2, 1, 2, 0, 1, 1, 0, 0, 0, 0} \, , \nn\\
F_{66} &= \epsilon^4(1-2\epsilon) \, m^4 \, I_{1, 2, 1, 0, 0, 0, 1, 1, 1, 0, 0, 1} \, , \nn\\
F_{67} &= \epsilon^5 \, (s-m^2) \, I_{1, 1, 0, 1, 1, -1, 1, 1, 0, 2, 0, 0} \, , \nn\\
F_{68} &= \epsilon^5 \, (s-m^2)^2 \, I_{1, 1, 0, 1, 1, 0, 1, 1, 0, 2, 0, 0} \, , \nn\\
F_{69} &= \epsilon^6 \, (s-m^2) \, I_{1, 1, 0, 1, 0, 0, 1, 1, 1, 0, 0, 1} \, , \nn\\
F_{70} &= \epsilon^6 \, (s-m^2)^2 \, I_{1, 1, 0, 1, 1, 0, 1, 1, 1, 1, -1, 1} \, , \nn\\
F_{71} &= \epsilon^6 \, (s-m^2) \, I_{1, 1, 0, 1, 1, -1, 1, 1, 1, 1, -1, 1}\nn \\&
+\frac{1}{12(1-2\epsilon)}(12F_2+6F_{3}+3F_4-2F_7+6F_9-18F_{14}+2F_{24}+12F_{25}) \, .
\label{eq:master}
\end{align}

Note that the choice of canonical basis is not unique.
We choose the basis in such a way that they have a uniform transcendentality, and,
more importantly, that it is easy to  determine their boundary conditions,
which will be discussed in the next section.

The differential equations for the above basis can  be expressed in the canonical form,
\bqa
\frac{\partial {\bf F}(x,\epsilon)}{\partial x} = \epsilon\left(\frac{\bf P}{x}+\frac{\bf Q}{x-1}\right){\bf F}(x,\epsilon).
\eqa
Here the variable $x$ is defined as $x\equiv \frac{s}{m^2}$.
The singular points at $x=0$ and $x=1$  correspond to the soft limit and threshold limit, respectively.
In analytic continuation, it is understood that $x\equiv \frac{s}{m^2}+i0$.
${\bf P}$ and ${\bf Q}$ are $71\times 71$ rational matrices,
of which the explicit forms are provided in the ancillary file.

\section{Boundary conditions and solutions of differential equations}
\label{sec3}

\subsection{Three-loop planar master integrals}
Before solving the differential equations shown in the previous section,
the boundary conditions must be determined.
We will make use of several properties of the basis integrals to achieve this goal.
First, we find that
the bases $\{F_1 \ldots F_4,F_{12},F_{14},F_{66}\}$ are single scale integrals and their results have been already known in the literature \cite{Henn:2016kjz},
\bqa
F_1&=&\frac{1}{8}\, , \nn\\
F_2&=&\frac{1}{8}+\epsilon^2\frac{\pi^2}{12}+\epsilon^3\zeta(3)+\epsilon^4\frac{4\pi^4}{45}+2\epsilon^5\frac{27\zeta(5)+\pi^2\zeta(3)}{3}
\nn\\&+ &\epsilon^6\left(\frac{229\pi^6}{1890}+4\zeta^2(3)\right) +{\cal O}(\epsilon^{7})  \, , \nn\\
F_3&=&-\frac{1}{6}-\epsilon^2\frac{\pi^2}{3}-\epsilon^3\frac{16\zeta(3)}{3}-\epsilon^4\frac{37\pi^4}{45}-16\epsilon^5\frac{39\zeta(5)+2\pi^2\zeta(3)}{3}
\nn\\ &-&\epsilon^6\left(\frac{2318\pi^6}{945}+\frac{256\zeta^2(3)}{3}\right) +{\cal O}(\epsilon^{7})  \, , \nn\\
F_4 &=& 1+8\epsilon^3\zeta(3)-\epsilon^4\frac{2\pi^4}{5}+144\epsilon^5\zeta(5)+\epsilon^6\left(-\frac{4\pi^6}{7}+32\zeta^2(3)\right)+{\cal O}(\epsilon^{7})\, ,\nn\\
F_{12} &=& -\frac{1}{6}-\epsilon^2\frac{\pi^2}{3}-\epsilon^3\frac{19\zeta(3)}{3}-\epsilon^4\frac{151\pi^4}{180}-\epsilon^5\left(\frac{38\pi^2\zeta(3)}{3}+215\zeta(5)\right)
\nn\\ &-&\epsilon^6\left(\frac{4729\pi^6}{1890}+\frac{361\zeta^2(3)}{3}\right)+{\cal O}(\epsilon^{7})\, , \nn\\
F_{14} &=&\frac{1}{12}-\epsilon^2\frac{\pi^2}{18}-\epsilon^4\frac{7\pi^4}{30}-\epsilon^5\left(\frac{8\pi^2\zeta(3)}{3}+34\zeta(5)\right)
\nn\\&-&\epsilon^6\left(\frac{4069\pi^6}{5760}+14\zeta^2(3)\right)+{\cal O}(\epsilon^{7})\, , \nn\\
F_{66} &=& -\epsilon^3\zeta(3)-\epsilon^4\frac{11\pi^4}{180}-38\epsilon^5\zeta(5)-\epsilon^6\left(\frac{431\pi^6}{1890}+18\zeta^2(3)\right)+{\cal O}(\epsilon^{7})\, .
\eqa

Second, we notice that all the master integrals are regular at  $x=0$, as expected, since the limit $s\to 0$ does not correspond to any physical pole.
This regular condition that all the integrals are finite in this limit can be employed
to create relations among the boundary conditions of different bases.
For instance, the differential equation for $F_{11}$ can be formulated as
\beq
\frac{\partial F_{11}}{\partial x}=\epsilon\left(\frac{3F_{11}-F_3}{x}-\frac{6F_{11}}{x-1}\right).
\eeq
Since $F_{11}$ is regular at $x\to 0$, its derivative should be also free of such a pole as $1/x$.
This means that on the right-hand side of the above equation, the coefficient of $1/x$ is vanishing in this limit, i.e.,
\beq
F_{11}|_{x=0}=\frac{1}{3} F_3|_{x=0}.
\eeq
%Note that $x\rightarrow 1$ is a singular point for $F_{11}$.

Third, the master integrals in $F_7,F_9,F_{18}$, and the integral $I_{0,1,0,2,0,2,0,1,1,0,0,0}$ in  $F_{21}$ do not contain any sub-topology,
and  we apply Mellin-Barnes integration method to calculate their boundary conditions at $x\rightarrow 0$.
The calculation is easy since the results are all expressed in terms of $\Gamma$-functions after using some functions
in the Mathematica packages \texttt{MB} \cite{Czakon:2005rk} and \texttt{AMBRE} \cite{Gluza:2007rt}.

We use $n_s$ to count the number of linear independent single scale integrals appearing in our calculation,
and denote as $n_0$ the number of linear independent master integrals
whose boundary conditions can be determined from the regular conditions at $x\rightarrow 0$.
The number of integrals whose boundary conditions can be calculated with Mellin-Barnes method is represented by $n_{mb}$.
We find that $n_s+n_0+n_{mb} < 71$, which means that
we are not able to determinate all the boundary conditions using the above three methods.

Note that more than half of the bases defined in eq. (\ref{eq:master}) contain a coefficient $(s-m^2)$.
Though $x\rightarrow 1$ may be a singular point for several integrals,
we find that there are some  bases that may be regular at $x\rightarrow 1$.
If this is the case, then  they are actually vanishing at $x\rightarrow 1$.
We use $n_1$ to represent the number of linear independent bases whose boundary conditions are vanishing at $x\rightarrow 1$.
It turns out that $n_s+n_0+n_{mb}+n_1 > 71$
so that we can determinate all the boundary conditions in simple ways.

As an example, we consider the topology $I_{0, n_2, n_3, n_4, 0, 0, n_7, n_8, n_9, 0, 0, 0}$ with $n_i>0$.
This topology has two master integrals. The canonical bases for this topology are chosen as
\bqa
F_{38} &=& \epsilon^5 \, (s-m^2) \, I_{0, 1, 1, 1, 0, 0, 1, 2, 1, 0, 0, 0} \, , \nn\\
F_{39} &=& \epsilon^4 \, m^2\, (s-m^2) \, I_{0, 1, 2, 1, 0, 0, 1, 2, 1, 0, 0, 0} \, ,
\eqa
and the corresponding differential equations for them are formulated as
\bqa
\frac{\partial F_{38}}{\partial x} &=& \epsilon\left(\frac{-4F_2-F_3+6F_{11}+2F_{19}-6(3F_{38}-2F_{39})}{6x}+\frac{2F_{38}}{x-1}\right)\, ,
\label{eq:F38}\\
\frac{\partial F_{39}}{\partial x} &=& \epsilon\left(\frac{-20F_2+4F_3+3F_{11}-12F_{16}+30F_{18}+16F_{19}-30(3F_{38}-2F_{39})}{12x}\right.\nn\\
& &\left.-2\frac{3F_{39}-4F_{38}}{x-1}\right)\, .
\eqa
We can derive two equations from the regular condition at $x\rightarrow 0$ for the differential equations of $F_{38}$ and $F_{39}$,
\bqa
-6(3F_{38}-2F_{39})|_{x=0}&=& (-4F_2-F_3+6F_{11}+2F_{19})|_{x=0}\, , \nn\\
-30(3F_{38}-2F_{39}) |_{x=0}&=& (-20F_2+4F_3+3F_{11}-12F_{16}+30F_{18}+16F_{19})|_{x=0}\, .
\label{eq:regular}
\eqa
However, one can readily see that these two equations are not linear independent,
hence only one of the boundary conditions for $F_{38},F_{39}$ at $x\rightarrow 0$ can be obtained from the regular conditions at $x=0$
\footnote{Since $F_{38}$ starts from weight four, the boundary conditions for the series of $F_{39}$ with weight less than four
can already be determined by eq.(\ref{eq:regular}). In the main text, we focus on the part with weight equal to or larger than four. }.
The other boundary condition must be obtained in another way.
Of course, one can use some standard methods, such as the numerical estimation or multi-fold Mellin-Barnes integration.
However we want to obtain analytic results and consider the Mellin-Barnes integration still complicated.
As a result, we adopt a guess-and-check method.
Observing the definitions of $F_{38},F_{39}$ in eq.(\ref{eq:master}),
we find that the denominator of $F_{39}$ has more powers than that of $F_{38}$,
indicating that $F_{38}$ may be less singular than $F_{39}$.
Therefore, we make a bold assumption that $F_{38}$ is regular at $x\to 1$.
Under this assumption,  we readily know that $F_{38}=0$ at $x= 1$
from its differential equation in eq.(\ref{eq:F38}).
After obtaining the boundary condition of $F_{38}$, we solve the differential equations for $F_{38}$.
Then we use the regular condition in eq.(\ref{eq:regular}) to obtain the boundary condition for $F_{39}$.
With the analytic results of $F_{38},F_{39}$ at hand,
we have checked them with the numerical results calculated by \texttt{FIESTA}
and found perfect agreement.
This confirms our assumption that $F_{38}$ is regular at $x\to 1$.
So far we have not figured out a principle to apply this method to more general integrals,
but it is efficient in practice, e.g. for the calculation of the nine-line master integral $F_{71}$.
For comparison, we notice that in the calculation of master integrals for massive form factors \cite{Henn:2016kjz}
all the boundary conditions except the single scale integrals can be obtained in the soft limit, i.e., $x\to 0$,
and the degeneracy problem in eq.(\ref{eq:regular}) does not happen there.

The methods described above can be used to determinate the boundary conditions for all bases in this work.
We list in table \ref{parameter} the specific method for each basis.
It is found that the bases $F_i, i=13,22,33,34,35,38,40,42,43,46,48,50,52,54,56,58,59,60,67,69,71$ are vanishing at $x=1$.
The bases $F_i,i=5,6,8,10,15\sim 17,20,24\sim 27,30\sim 32,44,62\sim 65$ are vanishing  at $x=0$.
Then the left non-vanishing boundary conditions are given by
\bqa
F_7|_{x=0} &=& -\frac{1}{2}-\epsilon^2\frac{\pi^2}{2}+5\epsilon^3\zeta(3)-\epsilon^4\frac{23\pi^4}{60}+\epsilon^5(27\zeta(5)+5\pi^2\zeta(3))\nn\\&-&\epsilon^6\left(\frac{361\pi^6}{1260}+25\zeta^2(3)\right)+{\cal O}(\epsilon^{7})\, ,\nn\\
F_9|_{x=0} &=& -\frac{1}{2}-\epsilon^2\frac{\pi^2}{6}+\epsilon^3\zeta(3)-\epsilon^4\frac{\pi^4}{20}+\epsilon^5\frac{(9\zeta(5)+\pi^2\zeta(3))}{3} \nn\\&-&\epsilon^6\left(\frac{61\pi^6}{3780}+\zeta^2(3)\right)+{\cal O}(\epsilon^{7})\, ,\nn\\
F_{11}|_{x=0} &=& \frac{1}{3}F_3\, ,\nn\\
F_{18}|_{x=0} &=&-\frac{1}{15}-\epsilon^2\frac{\pi^2}{10}-\epsilon^3\frac{14\zeta(3)}{15}-\epsilon^4\frac{33\pi^4}{200}-\epsilon^5\frac{7\pi^2\zeta(3)+134\zeta(5)}{5}
\nn \\ &-& \epsilon^6\left(\frac{2713\pi^6}{8400}+\frac{98\zeta^2(3)}{15}\right)+{\cal O}(\epsilon^{7})\, , \nn\\
F_{19}|_{x=0}&=&-5F_{18}|_{x=0}\, ,\nn\\
F_{21}|_{x=0} &=& -2\epsilon^3\zeta(3)-\epsilon^4\frac{\pi^4}{30}-2\epsilon^5(\pi^2\zeta(3)+7\zeta(5))
\nn\\&+& \epsilon^6\left(-\frac{41\pi^6}{630}+14\zeta^2(3)\right)+{\cal O}(\epsilon^{7})\, , \nn\\
F_{23}|_{x=0} &=&\frac{1}{24}-\epsilon^3\frac{5\zeta(3)}{3}-\epsilon^4\frac{2\pi^4}{15}-\epsilon^5(3\pi^2\zeta(3)+53\zeta(5))
\nn \\ &-& \epsilon^6\left(\frac{173\pi^6}{315}+\frac{62\zeta^2(3)}{3}\right)+{\cal O}(\epsilon^{7})\, , \nn\\
F_{28}|_{x=0} &=& \frac{1}{3}F_{12}|_{x=0}\, ,\nn\\
F_{29}|_{x=0} &=& -\frac{1}{3}F_{13}|_{x=0}\, ,\nn\\
F_{36}|_{x=0} &=&-\frac{7}{120}+\epsilon^2\frac{17\pi^2}{360}+\epsilon^3\frac{7\zeta(3)}{4}+\epsilon^4\frac{47\pi^4}{1800}
-\epsilon^5\left(\frac{11\pi^2\zeta(3)}{3}+\frac{7\zeta(5)}{20}\right)
\nn\\ &-&\epsilon^6\left(\frac{4361\pi^6}{8100}+\frac{69\zeta^2(3)}{2}\right)+{\cal O}(\epsilon^{7})\, , \nn\\
F_{37}|_{x=0} &=&-\frac{11}{240}-\epsilon^2\frac{\pi^2}{180}+\epsilon^3\frac{3\zeta(3)}{4}+\epsilon^4\frac{379\pi^4}{5400}
-\epsilon^5\left(\frac{4\pi^2\zeta(3)}{3}-\frac{1107\zeta(5)}{20}\right)
\nn\\ &+&\epsilon^6\left(\frac{901\pi^6}{4725}+\frac{21\zeta^2(3)}{2}\right)+{\cal O}(\epsilon^{7})\, , \nn\\
F_{39}|_{x=0} &=&-\epsilon^2\frac{\pi^2}{36}+\epsilon^4\frac{79\pi^4}{1080}+\epsilon^5\left(\frac{143\pi^2\zeta(3)}{18}+\frac{5\zeta(5)}{2}\right)
\nn\\ &+&\epsilon^6\left(\frac{18737\pi^6}{22680}+48\zeta^2(3)\right)+{\cal O}(\epsilon^{7})\, , \nn\\
F_{41}|_{x=0} &=&\frac{7}{180}-\epsilon^2\frac{7\pi^2}{270}-\epsilon^3\frac{89\zeta(3)}{45}-\epsilon^4\frac{139\pi^4}{900}-\epsilon^5\frac{353\pi^2\zeta(3)+8469\zeta(5)}{135}
\nn\\ &-&\epsilon^6\left(\frac{92077\pi^6}{170100}+\frac{2503\zeta^2(3)}{45}\right)+{\cal O}(\epsilon^{7})\, , \nn\\
F_{45}|_{x=0} &=&\epsilon^2\frac{\pi^2}{36}+2\epsilon^3\zeta(3)+\epsilon^4\frac{\pi^4}{20}
-\epsilon^5\left(\frac{11\pi^2\zeta(3)}{9}-55\zeta(5)\right)
\nn\\ &-&\epsilon^6\left(\frac{311\pi^6}{2835}+23\zeta^2(3)\right)+{\cal O}(\epsilon^{7})\, , \nn\\
F_{47}|_{x=0} &=&-\epsilon^2\frac{\pi^2}{36}+\epsilon^3\frac{\zeta(3)}{2}+\epsilon^4\frac{127\pi^4}{1080}
+\epsilon^5\left(\frac{191\pi^2\zeta(3)}{18}-\frac{25\zeta(5)}{2}\right)
\nn\\ &+&\epsilon^6\left(\frac{19847\pi^6}{22680}-\frac{7\zeta^2(3)}{2}\right)+{\cal O}(\epsilon^{7})\, , \nn\\
F_{49}|_{x=0} &=&-\epsilon^2\frac{\pi^2}{30}+\epsilon^3\frac{2\zeta(3)}{5}+\epsilon^4\frac{11\pi^4}{75}+\epsilon^5\frac{163\pi^2\zeta(3)}{15}
\nn\\ &+&\frac{52}{1575}\epsilon^6(26\pi^6+945\zeta^2(3))+{\cal O}(\epsilon^{7})\, , \nn\\
F_{51}|_{x=0} &=&-\frac{5}{48}+\epsilon^2\frac{\pi^2}{36}+\epsilon^3\frac{7\zeta(3)}{12}+\epsilon^4\frac{\pi^4}{360}
-\epsilon^5\left(\frac{14\pi^2\zeta(3)}{9}-\frac{41\zeta(5)}{4}\right)
\nn\\ &-&\epsilon^6\left(\frac{1297\pi^6}{22680}+\frac{17\zeta^2(3)}{6}\right)+{\cal O}(\epsilon^{7})\, , \nn\\
F_{53}|_{x=0} &=&-\epsilon^2\frac{\pi^2}{18}-\epsilon^3\frac{5\zeta(3)}{3}-\epsilon^4\frac{19\pi^4}{270}+\epsilon^5\frac{20}{9}(\pi^2\zeta(3)-12\zeta(5))
\nn\\ &+&\epsilon^6\left(\frac{113\pi^6}{756}+\frac{47\zeta^2(3)}{3}\right)+{\cal O}(\epsilon^{7})\, , \nn\\
F_{55}|_{x=0} &=&-\frac{1}{72}-\epsilon^2\frac{\pi^2}{27}-\epsilon^3\frac{11\zeta(3)}{6}-\epsilon^4\frac{29\pi^4}{240}
+\epsilon^5\left(\frac{59\pi^2\zeta(3)}{108}-\frac{123\zeta(5)}{2}\right)
\nn\\ &-&\epsilon^6\left(\frac{5993\pi^6}{22680}+\frac{571\zeta^2(3)}{36}\right)+{\cal O}(\epsilon^{7})\, , \nn\\
F_{57}|_{x=0} &=&-\frac{1}{48}-\epsilon^2\frac{5\pi^2}{72}-\epsilon^3\frac{13\zeta(3)}{6}-\epsilon^4\frac{49\pi^4}{540}
+\epsilon^5\left(\frac{46\pi^2\zeta(3)}{9}-66\zeta(5)\right)
\nn\\ &+&\epsilon^6\left(\frac{1189\pi^6}{11340}+\frac{94\zeta^2(3)}{3}\right)+{\cal O}(\epsilon^{7})\, , \nn\\
F_{61}|_{x=0} &=&\frac{1}{144}+\epsilon^2\frac{37\pi^2}{2160}+\epsilon^3\frac{67\zeta(3)}{120}+\epsilon^4\frac{239\pi^4}{7200}+\frac{7}{1080}\epsilon^5(17\pi^2\zeta(3)+2475\zeta(5))
\nn\\ &+&\epsilon^6\left(\frac{61291\pi^6}{680400}+\frac{2843\zeta^2(3)}{360}\right)+{\cal O}(\epsilon^{7})\, , \nn\\
F_{68}|_{x=0} &=&-\frac{1}{60}-\epsilon^2\frac{\pi^2}{36}-\epsilon^3\zeta(3)-\epsilon^4\frac{7\pi^4}{150}
+\epsilon^5\left(\frac{73\pi^2\zeta(3)}{45}-\frac{136\zeta(5)}{5}\right)
\nn\\ &+&\epsilon^6\left(\frac{193\pi^6}{22680}+\frac{29\zeta^2(3)}{3}\right)+{\cal O}(\epsilon^{7})\, , \nn\\
F_{70}|_{x=0} &=&\frac{1}{240}+\epsilon^2\frac{\pi^2}{108}+\epsilon^3\frac{19\zeta(3)}{45}+\epsilon^4\frac{3007\pi^4}{64800}+\epsilon^5\frac{443\pi^2\zeta(3)+6672\zeta(5)}{360}
\nn\\ &+&\epsilon^6\left(\frac{9931\pi^6}{51030}+\frac{9719\zeta^2(3)}{360}\right)+{\cal O}(\epsilon^{7})\, .
\eqa

\begin{table}[h]
\centering
\caption{The methods of determining boundary conditions}
\label{parameter}
\begin{tabular}{|c|c|}\hline\hline
Methods  &  Basis integrals \\
\hline\hline  Single scale &  $F_1 \sim F_4, F_{12},F_{14},F_{66}$ \\ \hline
Mellin-Barnes &  $F_7,F_9,F_{18},F_{21}$ \\\hline
Vanishing at $x=0$ & $F_5,F_6,F_8,F_{10},F_{15}\sim F_{17},F_{20}$,\\ & $F_{24}\sim F_{27}$,$F_{30}\sim F_{32},F_{44},F_{62}\sim F_{65}$   \\
\hline
Vanishing at $x=1$ & $F_{13},F_{22},F_{33}\sim F_{35},F_{38},F_{40},F_{42},F_{43},F_{46},F_{48}$ \\ & $F_{50},F_{52},F_{54},F_{56},F_{58}\sim F_{60},F_{67},F_{69},F_{71}$ \\ \hline
Regular conditions   & $F_{11},F_{19},F_{23},F_{28},F_{29},F_{36},F_{37},F_{39},F_{41},F_{45}$ \\ & $F_{47},F_{49},F_{51},F_{53},F_{55},F_{57},F_{61},F_{68},F_{70}$   \\ \hline\hline
\end{tabular}
\end{table}

By now, we have determined all the necessary boundary conditions.
The differential equations can readily be solved using the method shown in ref.\cite{Henn:2013pwa}.
The analytic results for $\{F_1\ldots F_{71}\}$ up to transcendental weight six are provided in an ancillary file.
As an example, we show the result for $F_{71}$,
\bqa
F_{71} &=& \epsilon^4\left(H_{0,1,0,1}(x)-H_{0,0,1,1}(x)+\frac{\pi^2}{6}H_{0,1}(x)-\frac{\pi^4}{30}\right)\nn\\
&+& \epsilon^5\bigg(-2H_{0,0,0,0,1}(x)-2H_{0,0,0,1,1}(x)-2H_{0,0,1,0,1}(x)-10H_{0,0,1,1,1}(x)\nn\\
&+& 2H_{0,1,0,1,1}(x)+6H_{0,1,1,0,1}(x)-\frac{\pi^2}{6}H_{0,0,1}(x)+\pi^2H_{0,1,1}(x)+2\zeta(3)H_{0,1}(x)\nn\\
&-& \frac{7\pi^2\zeta(3)}{6}-\zeta(5)\bigg)\nn\\
&+&\epsilon^6\bigg(-\Big(2\zeta(5)+\frac{\pi^2\zeta(3)}{3}\Big)H_{1}(x)+\frac{9\pi^4}{40}H_{0,1}(x)\nn\\
&+&\zeta(3)(-13H_{0,0,1}(x)+9H_{0,1,1}(x)-2H_{1,0,1}(x))-\pi^2\Big(-H_{0,0,0,1}(x)-\frac{5}{6}H_{0,0,1,1}(x)\nn\\
&+& H_{0,1,0,1}(x)+6H_{0,1,1,1}(x)+\frac{1}{3}H_{1,0,0,1}(x)\Big)-11H_{0,0,0,0,0,1}(x)-11H_{0,0,0,0,1,1}(x)\nn\\
&-&20H_{0,0,0,1,0,1}(x)-20H_{0,0,0,1,1,1}(x)-16H_{0,0,1,0,0,1}(x)-26H_{0,0,1,0,1,1,}(x)\nn\\
&-& 29H_{0,0,1,1,0,1}(x)-76H_{0,0,1,1,1,1}(x)-14H_{0,1,0,0,0,1}(x)-12H_{0,1,0,0,1,1}(x)\nn\\
&+&2H_{0,1,0,1,0,1}(x)-4H_{0,1,0,1,1,1}(x)+3H_{0,1,1,0,0,1}(x)+12H_{0,1,1,0,1,1}(x)\nn\\
&+&36H_{0,1,1,1,0,1}(x)+4H_{1,0,0,0,0,1}(x)+2H_{1,0,0,1,0,1}(x)+2H_{1,0,1,0,0,1}(x)\nn\\
&-&\frac{1219\pi^6}{15120}\bigg)+{\cal O}(\epsilon^{7} ),
\eqa
where $H_{a_1,a_2,...,a_n}(x)~(a_i\in\{0,\pm 1\})$ are harmonic polylogarithms defined in \cite{Remiddi:1999ew}.

All the analytic results have been checked with the numerical package \texttt{FIESTA} \cite{Smirnov:2015mct},
and good agreement has been found. For illustration, we consider the most complicated nine-line master integrals
in $F_{70}$ and $F_{71}$, i.e., $I_{1, 1, 0, 1, 1, 0, 1, 1, 1, 1, -1, 1}$ and $I_{1, 1, 0, 1, 1, -1, 1, 1, 1, 1, -1, 1}$,
of which the analytic results  at the kinematic point $(m= 1.0,s= -1.3)$ are
\bqa
I^{\rm analytic}_{1, 1, 0, 1, 1, 0, 1, 1, 1, 1, -1, 1} &=& \frac{0.00078765}{\epsilon^6}-\frac{0.00393624}{\epsilon^5}+\frac{0.0190587}{\epsilon^4}- \frac{0.0151068}{\epsilon^3}\nn\\ &+&\frac{0.290244}{\epsilon^2}+\frac{1.37654}{\epsilon}+4.82542,\nn\\
I^{\rm analytic}_{1, 1, 0, 1, 1, -1, 1, 1, 1, 1, -1, 1} &=& \frac{-6.69426}{\epsilon}-63.1207.\nn
\eqa
The results evaluated by \texttt{FIESTA} at the same kinematic point   are given by
\bqa
I^{\rm numeric}_{1, 1, 0, 1, 1, 0, 1, 1, 1, 1, -1,1} & = &\frac{0.000788}{\epsilon^6} - \frac{0.003936}{\epsilon^5} + \frac {0.019058\pm 0.000002} {\epsilon^4} - \frac{0.015109\pm 0.000035}{\epsilon^3} \nn \\ & + &\frac{0.290192\pm 0.000756} {\epsilon^2} + \frac{1.37606\pm 0.01581}{\epsilon} + 4.80886\pm 0.31758, \nn \\
I^{\rm numeric}_{1, 1, 0, 1, 1, -1, 1, 1, 1, 1, -1,1} & = &\frac{-6.69429\pm 0.00003}{\epsilon} - 63.1213\pm0 .0004, \nn
\eqa
where we have included the numerical uncertainties.
We can see that the numerical results obtained using the package \texttt{FIESTA}  agree well with  our analytic results. Note that it takes \texttt{FIESTA} two days on an eight-cores workstation to reach the above precision,
while it takes only a few seconds to evaluate analytic results by running the Mathematica package \texttt{HPL} \cite{Maitre:2005uu} on a single core.

\subsection{Two-loop non-planar master integrals}
\begin{figure}[h]
\begin{center}
\includegraphics[scale=0.5]{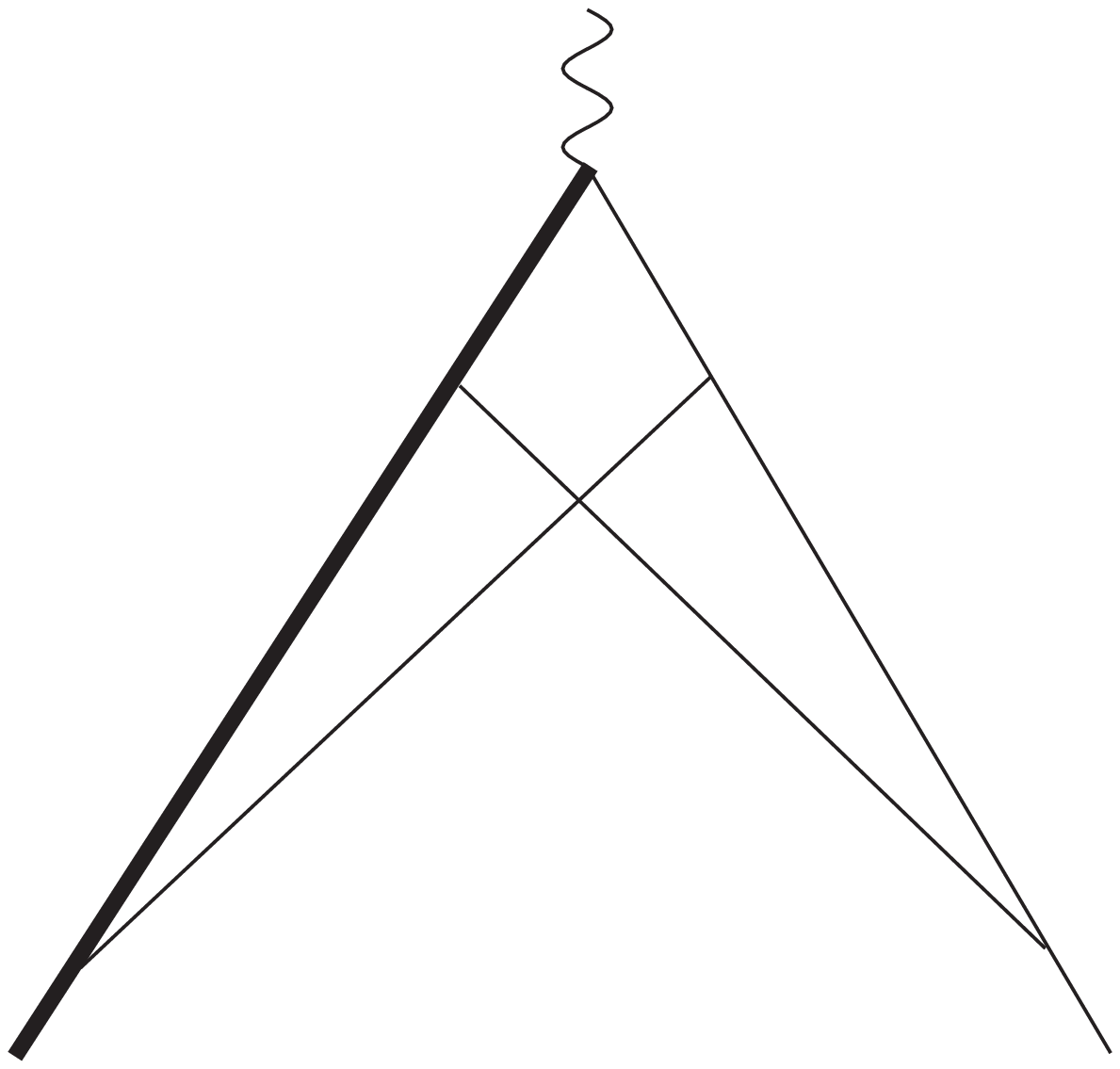}
\caption{The non-planar two-loop topology for the heavy-to-light form factor. The bold lines represent massive propagators, while the thick lines indicate massless propagators.}
\label{tnp}
\end{center}
\end{figure}

We have seen above that choosing a proper basis can efficiently simplify the determination of boundary conditions for master integrals.
This strategy is general and can also be used to calculate the non-planar master integrals.
For a proof-of-principle study, we consider the two-loop non-planar master integrals for the heavy-to-light form factor,
leaving the  results of three-loop non-planar master integrals  to a future work.
The master integrals shown in figure \ref{tnp} are represented generally by
\bqa
J_{n_1,\ldots,n_7}&=& \int\frac{{\mathcal D}^d k_1~{\mathcal D}^d k_2}{[-(k_1+p_1)^2+m^2]^{n_1}[-(k_2+p_1)^2+m^2]^{n_2}[-k_1^2]^{n_3}}\nn\\
& & \times \frac{1}{[-(k_2+p_2)^2]^{n_4}[-(k_1-k_2)^2]^{n_5}[-(k_2-k_1+p_2)^2]^{n_6}[-k_2^2]^{n_7}}.
\label{def2}
\eqa

The canonical basis is chosen as
\bqa
K_1 &=& \epsilon^2 \, J_{2, 2, 0, 0, 0, 0, 0} \, , \nn\\
K_2 &=& \epsilon^2 \, m^2 \, J_{0, 2, 2, 0, 1, 0, 0} \, , \nn\\
K_3 &=& \epsilon^2 \, s \, J_{2, 2, 0, 1, 0, 0, 0} \, , \nn\\
K_4 &=& \epsilon^2 \, s \, J_{0, 2, 2, 0, 0, 1, 0} \, , \nn\\
K_5 &=& \epsilon^2 \, (s-m^2) \, J_{0, 1, 2, 0, 0, 2, 0}-\frac{2m^2}{s} K_4 \, , \nn\\
K_6 &=& \epsilon^3 \, (s-m^2) \, J_{0, 1, 1, 1, 2, 0, 0} \, , \nn\\
K_7 &=& \epsilon^3 \, (s-m^2) \, J_{1, 2, 1, 0, 0, 1, 0} \, , \nn\\
K_8 &=& \epsilon^2 \, \frac{m^2\, s\, (s-m^2)}{s+m^2} \, J_{2, 2, 1, 0, 0, 1, 0}-\frac{m^2}{2(s+m^2)}(K_1-4K_4+K_5) \, , \nn\\
K_9 &=& \epsilon^3 \, (s-m^2) \, J_{2, 1, 0, 1, 0, 1, 0} \, , \nn\\
K_{10} &=& \epsilon^2 \, m^2\, (s-m^2) \, J_{3, 1, 0, 1, 0, 1, 0} \, , \nn\\
K_{11} &=& \epsilon^2 \, m^2(s-2m^2) J_{2, 2, 0, 1, 0, 1, 0}
+\frac{(s-2m^2)}{s-m^2} (2K_{10}-3K_9) \, , \nn\\
K_{12} &=& \epsilon^4 \, (s-m^2) \, J_{1, 1, 1, 1, 1, 0, 0} \, , \nn\\
K_{13} &=& \epsilon^3 \, (s-m^2)^2 \, J_{1, 1, 1, 2, 1, 0, 0} \, , \nn\\
K_{14} &=& \epsilon^4 \, (s-m^2) \, J_{1, 1, 1, 1, 1, 1, -1} \, .
\label{twonpb}
\eqa
The corresponding differential equations for the basis ${\bf K}=\{K_1,K_2,...,K_{14}\}$ are given by
\bqa
\frac{\partial {\bf K}(y,\epsilon)}{\partial y} =
\epsilon\left(\frac{\bf L}{y}+\frac{\bf M}{y-1}+\frac{\bf N}{y+1}\right){\bf K}(y,\epsilon)
\eqa
with $y\equiv \frac{s-m^2}{m^2}$.  $\bf L,M,N$ are rational matrices  provided in an ancillary file.
The relation between $y$ and the previous $x$ is $y=x-1$. We use $y$ as a letter in the above differential equations
so that the singularities appear at $-1,0,1$.
The singularities at $y=-1,0$ corresponds to the poles at $x=0,1$ in the planar diagrams.
The new singularity at $y=1$ is a pseudo-pole because it has no physical origin.
It appears only in the differential equations, but not in the final result.
Actually, this property can be used to derive the boundary conditions of some bases.
For example, since the loop integral is regular at $y=1$,
$K_{11}$ is equal to zero at $y=1$ due to the prefactor $(s-2m^2)$.

The determination of the other boundary conditions is similar to that in the three-loop planar case.
The single scale bases are
\bqa
K_1 &=& 1 \, ,\nn\\
K_2 &=& \frac{1}{4}+\epsilon^2\frac{\pi^2}{6}+2\epsilon^3\zeta(3)+\epsilon^4\frac{8\pi^4}{45}+{\cal O}(\epsilon^{5}) \, .
\eqa
The bases $K_3,K_4$ are vanishing at $y=-1$, and
$K_7,K_9,K_{10},K_{12},K_{14}$ are vanishing at $y=0$.
The other non-vanishing boundary conditions  are estimated to be
\bqa
K_5|_{y=-1} &=& -1 - \epsilon^2\frac{\pi^2}{3}+2\epsilon^3\zeta(3)-\epsilon^4\frac{\pi^4}{10}+{\cal O}(\epsilon^{5}) \, ,\nn\\
K_6|_{y=-1} &=& \frac{K_2}{2} \, , \nn\\
K_8|_{y=-1} &=& \epsilon^2\frac{\pi^2}{6} - \epsilon^3\zeta(3) + \epsilon^4\frac{\pi^4}{20} + {\cal O}(\epsilon^{5}) \, ,\nn\\
K_{13}|_{y=-1} &=& -\frac{5}{24} + \epsilon^2\frac{\pi^2}{18} + \epsilon^3\frac{7\zeta(3)}{6} + \epsilon^4\frac{\pi^4}{180} + {\cal O}(\epsilon^{5}) \, .
\eqa

With these boundary conditions, it is straightforward to solve the differential equations to obtain the
master integrals at general kinematics.
The analytic results of $\{K_1\ldots K_{14}\}$ are provided in an ancillary file along with this paper.

At the end of this section,
we explain a key point in our choice of the basis integrals for the six-line integral.
Usually, one would expect to choose $\epsilon^4 (s-m^2)^2 \, J_{1, 1, 1, 1, 1, 1, 0}$
as a canonical basis integral \cite{Bonciani:2008wf,Mastrolia:2017pfy}.
However, we could not determinate the boundary condition for this basis from the regular condition at $y=-1$,
and it has a logarithmic singularity at $y\rightarrow 0$.
One may also expect to use other ways to obtain the boundary conditions.
The integral at the boundary $y=-1$ has been calculated numerically in \cite{Bonciani:2008wf}.
In ref. \cite{Huber:2009se}, the author calculated the boundary condition for $J_{1, 1, 1, 1, 1, 1, 0}$ at $y=-1$
by computing several three-fold Mellin-Barnes integrals.
And in ref. \cite{Mastrolia:2017pfy}  the boundary condition for $J_{1, 1, 1, 1, 1, 1, 0}$ at  $y\rightarrow \infty$
has been calculated using the idea outlined in \cite{Dulat:2014mda}.

In our calculation we choose an alternative basis integral $K_{14} = \epsilon^4 \, (s-m^2) \, J_{1, 1, 1, 1, 1, 1, -1}$
instead of $\epsilon^4 \, (s-m^2)^2 \, J_{1, 1, 1, 1, 1, 1, 0}$.
Due to the numerator, the master integral $J_{1, 1, 1, 1, 1, 1, -1}$ is expected to be less divergent.
Therefore we use the guess-and-check method and assume that it is regular at $y=0$.
Then we see from the its differential equation that $K_{14}=0$ at $y\to 0$.
Under this assumption, we solve the differential equation and obtain the analytic result.
The correctness of our assumption is checked by the comparison with \texttt{FIESTA}.

In order to compare with the previous result in the literature,
we calculate  the result of $J_{1, 1, 1, 1, 1, 1, 0}$  from $J_{1, 1, 1, 1, 1, 1, -1}$ and other known bases by applying the IBP reduction.
The value of $\epsilon^4 \, (s-m^2)^2 \, J_{1, 1, 1, 1, 1, 1, 0}$ at $y=-1$ is determined to be
\bqa
\epsilon^4 \, (s-m^2)^2 \, J_{1, 1, 1, 1, 1, 1, 0}|_{y=-1}=\frac{1}{12}-\frac{7\pi^2}{72}\epsilon^2-\frac{89\zeta(3)}{12}\epsilon^3-\frac{71\pi^4}{120}\epsilon^4+ {\cal O}(\epsilon^{5}),
\eqa
which agrees with the result in ref. \cite{Huber:2009se}.

\section{Conclusions}
\label{sec4}

We have calculated analytically the color-planar three-loop master integrals for the heavy-to-light form factor,
which is necessary to provide a precision prediction for the heavy quark production and decay.
We make use of the differential equations to calculate the master integrals.
After choosing a canonical basis properly, the boundary conditions can be determined easily.
This is achieved by studying the pole structure of the master integrals carefully and
making use of the guess-and-check method.
As a result, the differential equations for the basis can be readily solved
and all master integrals are expressed in terms of harmonic polylogarithms.
The rational matrices of the differential equations and the analytic results of the master integrals  are all provided in ancillary files.
It would  be interesting to extend our method to the other diagrams such as the color suppressed
or non-planar three-loop master integrals
in order to calculate the full heavy-to-light form factor at the three-loop level.

\section*{Acknowledgments}
\noindent
The work of L.-B.C was supported by the National Natural Science Foundation of China (NSFC) under the grants 11747051 and 11805042.
The work of J.W was supported  by the BMBF project No. 05H15WOCAA and 05H18WOCA1.
J.W gratefully acknowledges the hospitality and partial support of the Mainz Institute for Theoretical Physics (MITP)
during the completion of this work.

%%%%%%%%%%%%%%%%%%%%%%%%%%%%%%%%%%%%%%%%%%%%

\bibliography{refs}
\bibliographystyle{JHEP}

%\begin{thebibliography}{99}

%\end{thebibliography}

%%%%%%%%%%%%%%%%%%%%%%%%%%%%%%%%%%%%%%%%%%%%

\end{document}